# Effect of Deep Learning Feature Inference Techniques on Respiratory Sounds

# Derin Öğrenme Öznitelik Çıkarma Tekniklerinin Solunum Sesleri Üzerindeki Etkisi

Osman Balli[1], Yakup Kutlu[1]
[1]Computer Engineering, Iskenderun Technical University, Hatay, Turkey
osman.balli50@gmail.com, yakup.kutlu@iste.edu.tr

*Abstract*—Analysis of respiratory sounds increases its importance every day. Many different methods are available in the analysis, and new techniques are continuing to be developed to further improve these methods. Features are extracted from audio signals and trained using different machine learning techniques. The use of deep learning, which is a different method and has increased in recent years, also shows its influence in this field. Deep learning techniques applied to the image of audio signals give good results and continue to be developed. In this study, image filters were applied to the values obtained from audio signals and the results of the features formed from this were examined in machine learning and deep learning techniques. Their results were compared with the results of methods that had previously achieved good results.

*Keywords—respiratory sounds; transfer learning.*

*Özetçe*—Solunum seslerinin analizi her geçen gün önemini daha da arttırmaktadır. Analizde birçok farklı yöntem mevcuttur ve bu yöntemleri daha da iyileştirmek için yeni teknikler geliştirilmeye devam edilmektedir. Genellikle ses sinyallerinden öznitelik çıkarımı yapılmakta ve farklı makine öğrenmesi teknikleri ile eğitim yapılmaktadır. Farklı bir yöntem olan ve son yıllarıda artış gösteren derin öğrenme kullanımı bu alanda da etkisini göstermektedir. Ses sinyallerinin görüntüsüne uygulanan derin öğrenme teknikleri iyi sonuçlar vermekte ve geliştirilmeye devam edilmektedir. Bu çalışmada, ses sinyallerinden elde edilen değerlere görüntü filtreleri uygulanmış ve bundan oluşan özelliklerin sonuçları makine öğrenimi ve derin öğrenme tekniklerinde incelenmiştir. Sonuçları daha önce iyi sonuçlar elde edilmiş yöntemlerin sonuçları ile kıyaslanmıştır.

*Anahtar Kelimeler—solunum sesleri; transfer öğrenimi.*

## I. INTRODUCTION

Many people experience respiratory disease. Many respiratory diseases are present, both seasonal and chronic. Although the diagnosis of these diseases is usually attempted with the help of a stethoscope, the procedure can be difficult for doctors. In order to overcome this challenge, artificial intelligence is especially helpful.

Artificial intelligence plays an important role in the diagnosis of respiratory diseases. It is possible to quickly diagnose diseases that cannot be fully diagnosed by doctors using artificial intelligence techniques. One of the important issues in this field is reliability. High performance of the system and high accuracies from the tested data will also improve the quality of the system to be created. Systems should be made more reliable, especially in the field of Health, in case minor errors can affect human life. A lot of work has been done and continues to be done in order to develop these systems.

There are different approaches to the analysis of respiratory sounds. Using machine learning methods using MFCC (Mel Frequency Cepstrum Coefficient) to an audio signal, applying deep learning methods by taking a spectrum image of an audio signal are the most common of them. Looking at the literature, many studies have been conducted in this field and good results have been obtained. The proposed Deep Extreme Learning Machines model with LU-based Autoencoder kernel has separated COPD (Chronic Obstructive Pulmonary Disease) and healthy subjects with high classification performance rates of 95.00%, 93.33% and 93.54% for accuracy, sensitivity and specificity [1], [2]. In another study, using DBN, they separated lung sounds from asthmatic patients and healthy subjects with classification performance rates of 91.35%, 89.54% and 83.94%, respectively, for overall accuracy, sensitivity and specificity [3]. 2 methods were used in a study conducted by Demir et al. The spectrum image of the audio signals was taken and attribute inference was made as the first method and performance was calculated with SVM (Support Vector Machines), the second method used transfer learning and softmax classifier. Accuracies of 65.60% and 63.09% respectively were achieved [4]. In a study conducted by Aykanat et al., MFCC was applied to the audio signal and the training results obtained with SVM were compared with the CNN result applied to the spectrum image [5]. In the study of Liu et al., CNN was applied by taking a spectrum image and obtained 81.16% headscarf [6]. Chamberlain et al. used attributes created with Denoising Autoencoder in the SVM classifier and achieved 86% and 74% for wheels and cracks [7]. The combined use of 3D-SODP quantization properties with DBN separated lung sounds from different COPD levels and separated them with high classification performance rates of 95.84%, 93.34% and 93.65% for accuracy, sensitivity and specificity, respectively [8]. A previous study of the database used in this study looked at the effect of window size on audio signals and was found that 93% accuracy when examining the results with machine learning techniques [9].

Attribute inference was made with CNN (Convolution Neural Network) in this study. But unlike traditional methods, audio signal data was used instead of images. The obtained qualifications were transferred to transfer learning and their performance was calculated using CNN, KNN (K Nearest Neighbor), SVM, DT (Decision Tree) methods.

## II. MATERIAL AND METHODS

### A. Database

In this study, Rocha et al. his respiratory sounds data set [10] was used. The data set used data from 103 people, including 35 healthy, 32 pneumonia and 36 COPD. The sampling frequency of each data is 44100. These records, which were at least 20 seconds, were divided into 1-second sections, and a total of 2055 data were created. Our data, which is 2055 units in total, is divided into 1849 training and 206 test data.

### B. Convolution Neural Network (CNN)

The use of deep learning has increased with the advent of high-performance hardware. One of the most commonly used methods is CNN. It is used in methods such as object recognition, signal processing, classification. One of the reasons CNN is used in so many different fields is that it gives successful results.

When we look at the CNN structure, we first encounter convolution and pooling layers. Convolution layer is the layer where filters and activation function are applied to the received image. The number and size of filters are adjusted on this layer. The resulting results are sent back to the convolution layer or the pooling layer. The Pooling layer combines useful features from the previous layer. This layer shows a decrease in data size. We can use Convolution and pooling layers to any number and extent we want and adjust their output. Data passing through these layers is transferred to the flatten layer to train a neural network model called the full connected layer. The Flatten layer translates this data into a one-dimensional matrix. In the neural network model, the weights are constantly updated and the learning process is performed by the back propagation method.

Full connected layer is used to train and classify data with the softmax classifier. The structure of CNN can be summarized in 3 stages. in the first stage, the data passes through the convolution and pooling layer, and in the next stage, a column vector is created. These contain the attributes of the first entries. In the final stage, weights are applied to estimate the labels and estimate values are determined [4]. The significance of the results is controlled by looking at the characteristics of the forecast values, such as accuracy, precision.

The audio data used in this study has a sampling frequency of 44100. This means that there are a total of 44100 attributes in each data we receive in a second. The attributes were converted to a 210x210 Matrix and added to the CNN model like a picture. Attributes obtained from the model along with the Model result were transferred to use different methods with transfer learning.

### C. Transfer Learning

Transfer Learning is the transfer of results from data previously trained for attribute and classification for use in another study [11]. Features or weights are derived from the large data set. Weights can be used in small data sets. Attributes, on the other hand, provide convenience in applying different methods. No need for repeat convolution and pooling operations, saving time and ease of operation.

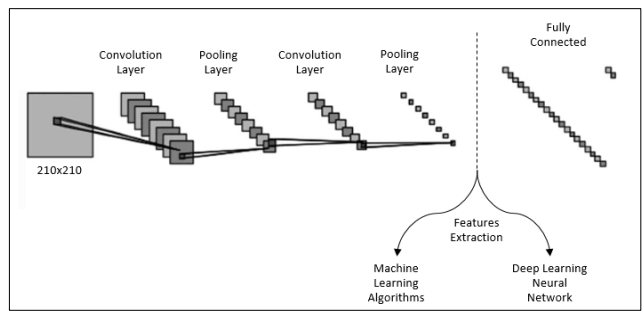

Figure 1: Transfer Learning Shema

Our data prepared as 210x210, convolution and pooling operations were applied as in Figure 1. After the last pooling layer, the features are taken and transferred for use in Machine Learning Algorithms. Here, while our data had 44100 attributes at first, 7744 attributes were obtained in the last case. Reducing attributes will speed up operations in machine learning methods. Results from Support Vector Machines, K Nearest Neighbor, Decision Tree methods and Fully Connected layer were compared.

### D. K Nearest Neighbor

The K Nearest Neighbor (KNN) algorithm is one of the simplest and most common learning methods. The distance of each data contained in the Test set to each data contained in the training set is calculated. For each data in the Test set, the nearest K sample is selected from the training set. Test data is assigned to the class that contains these values, the majority of which class belongs to the selected K sample.

In this study, the 3 nearest neighbors were used.

*E. Support Vector Machines*

SVM (Support Vector Machine), a machine learning method, is a supervised learning tool that can be used in regression or classification. SVM creates a model at the training stage, and data located in points in space is divided into classes by Hyperplane. The hyperplane varies in position by looking at the distance of the data. The number of hyperplanes is determined by the number of classes created. When new data is added, It looks at its position in space and looks at which plane it remains in and is labeled to that class. Nonlinear classification is effectively used thanks to SVM [12].

*F. Decision Tree*

It is one of the most commonly used supervised learning algorithms. A decision tree creates decision paths that can be created for a possible result. They can be used in regression or classification. A decision tree is extracted by selecting a feature (column) in the data set that we have and associating its branches on other fields. A decision tree shows a test of each internal node on an attribute. Each branch represents a result of the test and is a flow chart similar to the tree structure in which each leaf node contains a class label. [13].

The decision tree consists of 2 stages: growth and pruning. According to the criteria set in the first stage, the growth process continues until the data is in the same class or stage. At the pruning stage, it generalizes the tree, eliminating errors. The accuracy of the pruning stages increases [13].

### III. RESULTS AND DISCUSSION

In the study, training and test results obtained from CNN and machine learning methods were compared. CNN's epoch number was set at 40. Training accuracy 96% test accuracy 86% was achieved. The highest training accuracy was found to be 100% in the decision tree, while the lowest test accuracy was also found to be 76% in the decision tree. In order to understand the reliability of the study results, a confusion matrix was created in all trainings. Precision, recall and f1 score values were calculated with confusion matrix. F1 score is a measure used very often when comparing classification algorithms. It is the harmonic mean of precision and recall values. For this reason, it is a good indicator of accuracy. Looking at the results, f1 score was highest with 86% and CNN was lowest with 76%. When we compare all the parameters, it seems that the most reliable results were obtained in CNN, although we achieved the highest training accuracy in the Decision Tree.

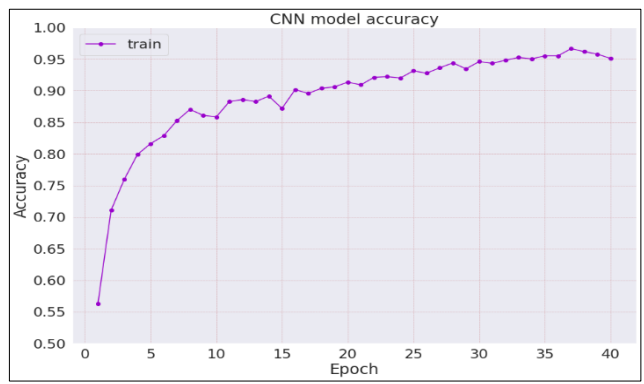

Figure 2: Accuracy versus Epoch for CNN Model

| Method | Train Accuracies | Test Accuracies | Precious | Recall | f1 score |
|---|---|---|---|---|---|
| CNN | 0.966 | 0.869 | 0.869 | 0.869 | 0.869 |
| KNN | 0.895 | 0.801 | 0.814 | 0.801 | 0.797 |
| SVM | 0.965 | 0.85 | 0.848 | 0.85 | 0.847 |
| DT | 1 | 0.762 | 0.759 | 0.762 | 0.761 |

Table 1: The other evaluation criteria for the proposed method

### REFERENCES

[1] Altan, G., Kutlu, Y., Pekmezci, A. & Yayık, A. (2018). Diagnosis of Chronic Obstructive Pulmonary Disease using Deep Extreme Learning Machines with LU Autoencoder Kernel. 7th International Conference on Advanced Technologies (ICAT'18). 618-622.

[2] Altan, G., Kutlu, Y., Garbi, Y., Pekmezci, A.Ö. & Nural, S., (2017). Multimedia Respiratory Database (RespiratoryDatabase@TR): Auscultation Sounds and Chest X-rays. NESciences, 2017, 2 (3): 59-72.

[3] Altan, G., Kutlu, Y., Pekmezci, A. & Nural, S. (2018). Asthma Analysis using Deep Learning. 7th International Conference on Advanced Technologies (ICAT'18).

[4] Demir, F., Sengur, A. & Bajaj, V. (2020). Convolutional neural networks based efficient approach for classification of lung diseases. Health Inf Sci Syst 8, 4.

[5] Aykanat, M., Kılıç, Ö., Kurt, B., & Saryal, S. (2017). Classification of lung sounds using convolutional neural networks. EURASIP Journal on Image and Video Processing, 2017(1).

[6] Liu, R., Cai, S., Zhang, K., & Hu, N. (2019). Detection of Adventitious Respiratory Sounds based on Convolutional Neural Network. 2019 International Conference on Intelligent Informatics and Biomedical Sciences (ICIIBMS).

[7] Chamberlain, D., Kodgule, R., Ganelin, D., Miglani, V., & Fletcher, R. R. (2016). Application of semi-supervised deep learning to lung sound analysis. 2016 38th Annual International Conference of the IEEE Engineering in Medicine and Biology Society (EMBC).

[8]  Altan, G., Kutlu, Y., Pekmezci, A. Ö., & Nural, S. (2018). Deep learning with 3D-second order difference plot on respiratory sounds. Biomedical Signal Processing and Control, 45, 58–69.

[9]  Balli, O., Kutlu, Y., (2019). Effect of Window Size for Detection of Abnormalities in Respiratory Sounds. Natural and Engineering Sciences 4(3):124-129.

[10] Rocha, B. M., Filos, D., Mendes, L., Vogiatzis, I., Perantoni, E., Kaimakais, E., Maglaveras, N. (2017). ICBHI 2017 Challenge Respiratory Sound Database. Retrieved from ICBHI Challenge: https://bhichallenge.med.auth.gr/

[11] Pan, S. J., & Yang, Q. (2010). A Survey on Transfer Learning. IEEE Transactions on Knowledge and Data Engineering, 22(10), 1345–1359.

[12] Palaniappan, R., & Sundaraj, K. (2013). Respiratory sound classification using cepstral features and support vector machine. 2013 IEEE Recent Advances in Intelligent Computational Systems (RAICS).

[13] Lavanya, D., & Rani, K. U. (2011). Performance Evaluation of Decision Tree Classifiers on Medical Datasets. International Journal of Computer Applications, 26(4), 1–4.